# A Second Source of Repeating Fast Radio Bursts

The CHIME/FRB Collaboration: M. Amiri[1], K. Bandura[2,3], M. Bhardwaj[4,5], P. Boubel[4,5], M. M. Boyce[6], P. J. Boyle[4,5], C. Brar[4,5], M. Burhanpurkar[7], T. Cassanelli[8,9], P. Chawla[4,5], J. F. Cliche[4,5], D. Cubranic[1], M. Deng[1], N. Denman[9,8], M. Dobbs[4,5], M. Fandino[1], E. Fonseca[4,5], B. M. Gaensler[9,8], A. J. Gilbert[4,5], A. Gill[8,9], U. Giri[10,11], D. C. Good[1], M. Halpern[1], D. S. Hanna[4,5], A. S. Hill[1,12,13], G. Hinshaw[1], C. Höfer[1], A. Josephy[4,5], V. M. Kaspi[4,5], T. L. Landecker[12], D. A. Lang[10,11], H.-H. Lin[14], K. W. Masui[15,16], R. Mckinven[9,8], J. Mena-Parra[4,5,15], M. Merryfield[4,5], D. Michilli[4,5], N. Milutinovic[12,1], C. Moatti[4,5], A. Naidu[4,5], L. B. Newburgh[17], C. Ng[9,*], C. Patel[4,5], U.-L. Pen[14], T. Pinsonneault-Marotte[1], Z. Pleunis[4,5], M. Rafiei-Ravandi[10], M. Rahman[9], S. M. Ransom[18], A. Renard[8], P. Scholz[12], J. R. Shaw[1,12], S. R. Siegel[4,5], K. M. Smith[10], I. H. Stairs[1], S. P. Tendulkar[4,5], I. Tretyakov[19,8], K. Vanderlinde[8,9], P. Yadav[1]

---

\* Corresponding Author
[1] *Department of Physics and Astronomy, University of British Columbia, 6224 Agricultural Road, Vancouver, BC V6T 1Z1, Canada*
[2] *CSEE, West Virginia University, Morgantown, WV 26505, USA*
[3] *Center for Gravitational Waves and Cosmology, West Virginia University, Morgantown, WV 26505, USA*
[4] *Department of Physics, McGill University, 3600 rue University, Montréal, QC H3A 2T8, Canada*
[5] *McGill Space Institute, McGill University, 3550 rue University, Montréal, QC H3A 2A7, Canada*
[6] *Department of Physics and Astronomy, University of Manitoba, 301 Allen Building, 30A Sifton Road, Winnipeg, MB R3T 2N2, Canada*
[7] *Harvard University, Cambridge, MA 02138, USA*
[8] *Department of Astronomy and Astrophysics, University of Toronto, 50 St. George Street, Toronto, ON M5S 3H4, Canada*
[9] *Dunlap Institute for Astronomy and Astrophysics, University of Toronto, 50 St. George Street, Toronto, ON M5S 3H4, Canada*
[10] *Perimeter Institute for Theoretical Physics, 31 Caroline Street N, Waterloo, ON N2L 2Y5, Canada*
[11] *Department of Physics and Astronomy, University of Waterloo, Waterloo, ON N2L 3G1, Canada*
[12] *Dominion Radio Astrophysical Observatory, Herzberg Astronomy & Astrophysics Research Centre, National Research Council of Canada, P.O. Box 248, Penticton, BC V2A 6J9, Canada*
[13] *Space Science Institute, Boulder, CO 80301 USA*
[14] *Canadian Institute for Theoretical Astrophysics, 60 St. George Street, Toronto, ON M5S 3H8, Canada*
[15] *MIT Kavli Institute for Astrophysics and Space Research, Massachusetts Institute of Technology, 77 Massachusetts Ave, Cambridge, MA 02139, USA*
[16] *Department of Physics, Massachusetts Institute of Technology, Cambridge, 77 Massachusetts Ave, MA 02139, USA*
[17] *Department of Physics, Yale University, New Haven, CT 06520, USA*
[18] *National Radio Astronomy Observatory, 520 Edgemont Road, Charlottesville, VA 22903, USA*
[19] *Department of Physics, University of Toronto, 60 St. George Street, Toronto, ON M5S 3H4, Canada*

**The discovery of a repeating Fast Radio Burst (FRB) source[1,2], FRB 121102, eliminated models involving cataclysmic events for this source. No other repeating FRB has yet been detected in spite of many recent FRB discoveries and follow-ups[3–5], suggesting repeaters may be rare in the FRB population. Here we report the detection of six repeat bursts from FRB 180814.J0422+73, one of the 13 FRBs detected[6] by the Canadian Hydrogen Intensity Mapping Experiment (CHIME) FRB project[7] during its pre-commissioning phase in July and August 2018. These repeat bursts are consistent with originating from a single position on the sky, with the same dispersion measure (DM), ~189 pc cm$^{-3}$. This DM is approximately twice the expected Milky Way column density, and implies an upper limit on the source redshift of 0.1, at least a factor of ~2 closer than FRB 121102[8]. In some of the repeat bursts, we observe sub-pulse frequency structure, drifting, and spectral variation reminiscent of that seen in FRB 121102[9,10], suggesting similar emission mechanisms and/or propagation effects. This second repeater, found among the first few CHIME/FRB discoveries, suggests that there exists – and that CHIME/FRB and other wide-field, sensitive radio telescopes will find – a substantial population of repeating FRBs.**

FRB 180814.J0422+73 was discovered by the CHIME/FRB project[6,7] in August 2018 when the FRB instrument was in a pre-commissioning phase. CHIME is a transit telescope and has an effective ~250 sq. deg instantaneous field-of-view (FOV). During the remainder of the Summer pre-commissioning, and during September and October commissioning periods, the position of this source was observed semi-regularly, with daily exposures (when not interrupted by commissioning activities) of ~36 min. The large exposure is facilitated by the source's high declination, which allows it to fall in the telescope's primary beam twice per day, 12-hour apart (on either side of the North Celestial Pole) and for ~18 min each transit.

Four additional bursts at the same DM were detected by CHIME/FRB in September (see Table 1) at sky positions consistent with the August event. As a detailed characterization of our beam shape requires a long-duration observation campaign that is in its early stages, we rely on analytically estimated formed beam shapes for the FFT beamforming[7,11] and a primary beam estimated from the CHIME Pathfinder telescope[12]. Fig. 1 shows the estimated source positions for five events in R.A. and Dec., along with our estimate of the true source position obtained by combining all five (see Methods). Our best position estimate is (J2000) R.A. $04^h22^m22^s$, Dec. +73°40′, with a 99% confidence uncertainty of ±4′ in R.A. and ±10′ in Dec.

A 6th event was detected on October 28 using the CHIME/Pulsar instrument (see Methods), which coherently dedisperses up to 10 formed and tracking beams at software-commanded sky positions for a specified DM. At the time of the burst, six CHIME/Pulsar beams were observing a grid covering the estimated position (see Methods) and four of the six beams detected the burst. Despite a pointing error in the grid centre, the CHIME/Pulsar instrument made its strongest detection of the burst in the beam closest to the best CHIME/FRB-derived position. The burst occurred in a side lobe of a CHIME/FRB formed beam and was initially assigned an incorrect R.A. and not classified as extragalactic. As the CHIME/Pulsar data has higher time resolution, we show that in Fig. 2.

We have searched for repeat bursts from the other 12 sources discovered during the pre-commissioning phase[6], by looking for events of similar DM when their best-estimated position was in the main lobes of the formed beams. We found no events exceeding our SNR threshold of 10. Each of the 12 was subject to a different exposure and sensitivity; two have higher declinations, hence more exposure, than for FRB 180814.J0422+73, so likely have significantly lower observed repeat rates, if they are repeaters. A detailed discussion of this will be presented elsewhere.

The automated pipeline[7] recorded raw intensity data to disk for all CHIME/FRB repeat events from FRB 180814.J0422+73 except for the burst on September 6, for which the system failed to record to disk. The events with intensity data allows us to examine their dynamic spectra (see Fig. 2) and measure refined burst parameters (see Table 1). The September 6 event has only metadata determined by the automated pipeline and therefore we have only coarse estimates of its properties; however, these are sufficient to verify that it was from the same source. Polarimetry of FRB 121102 revealed one of the highest rotation measures ever seen[9], an important clue about the source environment. No polarization information was available for the events reported here, however, functionality to record data with a higher time resolution and polarization information is currently being deployed for both the CHIME/FRB and CHIME/Pulsar systems.

The four events with raw intensity data show complexities in their burst profiles (Fig. 2). The events on September 17 and October 28 exhibit multiple spectro-temporal structures, reminiscent of bursts from FRB 121102[9,10,13] and FRB 170827[14]. Because of this structure, the burst fitting algorithm used in our companion paper[6], which assumes a simple Gaussian underlying profile, is not optimal for the bursts considered here. We therefore used a fitting routine that optimizes burst temporal structure[10], searching over a range of DMs from 186 to 206 pc cm$^{-3}$ (see Methods). Results for individual bursts are presented in Table 1. Assuming the structure is intrinsic to the FRB, the DM that optimizes structure across all five bursts is 189.4±0.4 pc cm$^{-3}$. We see no evidence for monotonic variation in the DMs.

In the dedispersed and summed time series for the September 17 and October 28 bursts, we identify 3 to 5 possible burst components above the median off-pulse intensity. Bursts have widths of a few milliseconds and, in events with sub-structure, peak separations of the same order. The sub-pulses appear to drift downwards in frequency as time progresses. To quantify this, we fit multiple 2D Gaussians in observing frequency and time to the bursts (see Methods) and infer frequency drift rates of −6.4±0.7 and −1.3±0.3 MHz ms$^{-1}$, for the September 17 and October 28 bursts, respectively (see Methods, Extended Data Table 1 and Extended Data Fig. 1). The measurement is robust against correcting for a frequency-dependent sensitivity model of the primary and formed beams. Note that burst structure shorter than ~1 ms is unresolved in our data. Similar sub-pulse frequency drifts have been seen in FRB 121102, also downwards, however at a much lower drift rate[9,10,13]. Interestingly, comparing drift rates from FRB 180814.J0422+73 with those for FRB 121102 shows they scale approximately linearly with frequency, although our time resolution makes us insensitive to drift rates much shorter than we have measured. The existence of such drifting and their same directionality in both known repeating sources is suggestive of a common physical origin, either intrinsic to the source, or due to propagation effects such as plasma lensing[15,16], although the latter should result in drift rates in both directions. The downward frequency drifts and the intermittent nature of the emission is reminiscent of type II solar radio bursts, which are coherent radio emissions from shock accelerated plasmas[17]. A related mechanism in a shocked, highly relativistic plasma outside a compact object might explain the behaviour seen in the FRB's bursts.

As our burst spectra are not yet calibrated for their frequency-dependent responses, we cannot infer the underlying source spectra. Systematic response differences are expected between the two different daily transits which are at different elevation angles, and order unity response variations across our bandwidth are expected for any elevation. However, from the large apparent spectral differences among bursts in the same transit, and noting that our localization analysis (Fig. 1) indicates no event was detected in a sidelobe, we infer qualitatively that burst spectra vary strongly. This phenomenon was also seen in the original repeater FRB 121102[1,2].

Throughout our pre-commissioning phase, the telescope and CHIME/FRB instrument were operated intermittently without a simple schedule and with varying levels of sensitivity. For this reason calculating the on-sky exposure to any position is presently challenging. Nevertheless, we have calculated an approximate exposure for the sky position of FRB 180814.J0422+73, and found (see Methods) that in total, it was approximately 23 hours from July 25 through November 4, where exposure is defined by the best estimate of the source's coordinates being within the telescope's primary beam. However, this exposure did not have uniform sensitivity, mainly because of the different transits: we estimate (see Methods) that the upper transit (14 hours of exposure), is approximately four times more sensitive than the lower transit (9 hours of exposure), due to the primary beam response. Even at the same transit, during this commissioning phase, different calibration strategies were invoked at different times. Thus, it is not possible to reconstruct the sensitivity as a function of time accurately. However, given that we detected 3 events above a fluence limit of 13 Jy ms in our band in 14 hours of exposure, we place a lower limit on the source average burst rate during these observations of 0.09 $hr^{-1}$ above this fluence at 95% confidence.

The DM due to the Milky Way disk along this line of sight is 87 pc $cm^{-3}$ in the NE2001 model[18] of the Galactic electron density distribution, and 100 pc $cm^{-3}$ in the YMW16 model[19]. We have verified there are no catalogued Galactic foregrounds[20,21] that could provide an anomalous DM contribution. This implies a DM excess over the Galactic value of ~89–102 pc $cm^{-3}$ or ~59–72 pc $cm^{-3}$, if we include a DM contribution from the halo[22] of the Milky Way of 30 pc $cm^{-3}$. Assuming all the excess DM is due to the intergalactic medium[23], we place an upper limit on the source redshift of $z < 0.11$. This corresponds to a distance of ~500 Mpc, approximately half the distance of FRB 121102.

The true redshift and distance are expected to be lower, since the host likely contributes to the DM. We have examined archival optical images of the field using the Panoramic Survey Telescope and Rapid Response System (PanSTARRS)[24] data and we find no obvious candidate host galaxy based on photometry or morphology (see Methods).

We searched a field of view of 0.04 $deg^2$ for possible persistent radio counterparts to FRB180814.J0422+73, like that seen for FRB 121102. An FRB 121102-like persistent radio source at the redshift $z = 0.09$ would have flux density ~0.72 mJy at 3 GHz, which should be detectable at $> 5\sigma$ significance in the Very Large Array (VLA) Sky Survey (VLASS)[25]. There are two data sets with relevant sensitivity: 1.4-GHz observations from the NRAO VLA Sky Survey (NVSS)[26], conducted in January 1994 down to an RMS of 0.48 mJy/beam, and an observation spanning 13 GHz from the VLASS, conducted in September 2017 down to an RMS of 0.13 mJy/beam. There are five NVSS sources detected in the field, with peak fluxes ranging from 2 to 60 mJy/beam. We detect only four of these sources in VLASS with peak flux $> 5\sigma$ significance, each of which is spectrally consistent with being a radio galaxy, unlike the flat-spectrum radio source co-located with FRB 121102. The fifth NVSS source, NVSS J042149+734107, has an NVSS peak flux of 6.4 mJy/beam (7.6 mJy integrated flux), but has an upper limit of 0.33 mJy/beam in VLASS. The source is plausibly an extended radio galaxy that is resolved out by the high angular resolution of VLASS (see Methods). Alternatively, it could have dimmed by a factor of $> 11$ over two decades, or have an unusually steep spectral index $\alpha < 3.9$.

Until now, the possibility that FRB 121102 was a unique – or at least a highly unusual – source was genuine, given the absence of detections of repetition in the >60 known FRBs, in spite of substantial follow-up observations[5,27]. CHIME/FRB, with its wide field and high sensitivity, was predicted to find a large population of repeaters, if they exist[28]. That one of our first 13 events repeats suggests there indeed exists a substantial population. The similarities between FRBs 121102 and 180814.J0422+73, namely their variable spectra, their substructure and their sub-components downward frequency drifts, may be coincidental, but may also provide hints about what distinguishes them from a putative non-repeating class, and could help observers identify likely repeaters among single-burst detections.

**Acknowledgements** We are grateful for the warm reception and skillful help we have received from the Dominion Radio Astrophysical Observatory, operated by the National Research Council Canada. The CHIME/FRB Project is funded by a grant from the Canada Foundation for Innovation 2015 Innovation Fund (Project 33213), as well as by the Provinces of British Columbia and Québec, and by the Dunlap Institute for Astronomy and Astrophysics at the University of Toronto. Additional support was provided by the Canadian Institute for Advanced Research (CIFAR), McGill University and the McGill Space Institute via the Trottier Family Foundation, and the University of British Columbia. The Dunlap Institute is funded by an endowment established by the David Dunlap family and the University of Toronto. Research at Perimeter Institute is supported by the Government of Canada through Industry Canada and by the Province of Ontario through the Ministry of Research & Innovation. The National Radio Astronomy Observatory is a facility of the National Science Foundation operated under cooperative agreement by Associated Universities, Inc. P.C. is supported by an FRQNT Doctoral Research Award and a Mitacs Globalink Graduate Fellowship. M.D. acknowledges support from the Canadian Institute for Advanced Research (CIFAR), NSERC Discovery and Accelerator Grants, and from FRQNT Centre de Recherche en Astrophysique du Québec (CRAQ). B.M.G. acknowledges the support of the Natural Sciences and Engineering Research Council of Canada (NSERC) through grant RGPIN-2015-05948, and the Canada Research Chairs program. A.S.H. is partly supported by the Dunlap Institute. V.M.K. holds the Lorne Trottier Chair in Astrophysics & Cosmology and a Canada Research Chair and receives support from an NSERC Discovery Grant and Herzberg Award, from an R. Howard Webster Foundation Fellowship from CIFAR, and CRAQ. C.M. is supported by a NSERC Undergraduate Research Award. J.M.-P. is supported by the MIT Kavli Fellowship in Astrophysics and a FRQNT postdoctoral research scholarship. M.M. is supported by a NSERC Canada Graduate Scholarship. Z.P. is supported by a Schulich Graduate Fellowship. S.M.R. is a CIFAR Senior Fellow and is supported by the NSF Physics Frontiers Center award 1430284. P.S. is supported by a DRAO Covington Fellowship from the National Research Council Canada. FRB research at UBC is supported by an NSERC Discovery Grant and by CIFAR.

**Author Information**

Reprints and permissions information is available at www.nature.com/reprints

The authors declare that they have no competing financial interests.

Correspondence and requests for materials should be addressed to C. Ng (email: cherry.ng@dunlap.utoronto.ca).

| Date | MJD (topocentric) | MJD (barycentric) | Transit | $t_{samp}$ (ms) | $DM_{struct}$ (pc cm$^{-3}$) | $DM_{SNR}$ (pc cm$^{-3}$) | Boxcar width (ms) | SNR | Fluence (Jy ms) |
|---|---|---|---|---|---|---|---|---|---|
| 180814[a] | 58344.61791692 | 58344.61702927 | upper | 0.983 | 189.0(1) | 190(4) | 7.9 | 9.8 | 21(15) |
| 180906[b] | 58367.05402060 | 58367.05573581 | lower | … | … | *191(3)* | *3.9* | *11* | … |
| 180911 | 58372.54113117 | 58372.54334254 | upper | 0.983 | 189.8(9) | 190(5) | 7.9 | 12 | 3.4(3) |
| 180917 | 58378.03235369 | 58378.03504878 | lower | 1.966 | 189.5(1) | 199(7) | 63 | 22 | 66(31) |
| 180919 | 58380.52510580 | 58380.52801573 | upper | 0.983 | 190.0(1) | 190(4) | 16 | 10 | 12(6) |
| 181028[c] | 58419.42536432 | 58419.43090504 | upper | 0.328 | 188.9(5) | 193(6) | 42 | 18.7 | … |

**Table 1: Properties of the detected bursts.** Properties of the five CHIME/FRB and one CHIME/Pulsar events from FRB 180814.J0422+73. Topocentric and barycentric arrival times are referred to 600 MHz and have uncertainty of ~1 ms. Transit "upper" refers to the main source transit at high elevation and "lower" refers to the secondary source transit at lower elevation. $t_{samp}$ is the time resolution of the saved data. $DM_{struct}$ is the DM determined for the source when optimizing for sub-pulse structure, while $DM_{SNR}$ is when maximizing SNR (see Methods). SNR and boxcar width are measured by convolving the time series dedispersed to DM=189.4 pc cm$^{-3}$ of each burst with a range of 1D boxcar kernels that are normalized by the square root of their widths and selecting the widths that result in the highest SNR (thereby neglecting the presence of burst sub-components). Note the difference in definition of width compared to that in our companion paper[6]. Values in italics are from metadata only as for that event, no raw intensity data were saved. Fluence calculations use between 178–216 MHz bandwidth and are described in the Methods section. Table note *a* refers to the discovery detection described in our companion paper[6]. Table note *b* indicates the event for which raw intensity data were not saved. Table note *c* indicates the event detected by the CHIME/Pulsar instrument, for which calibrated fluence measurements are not yet available.

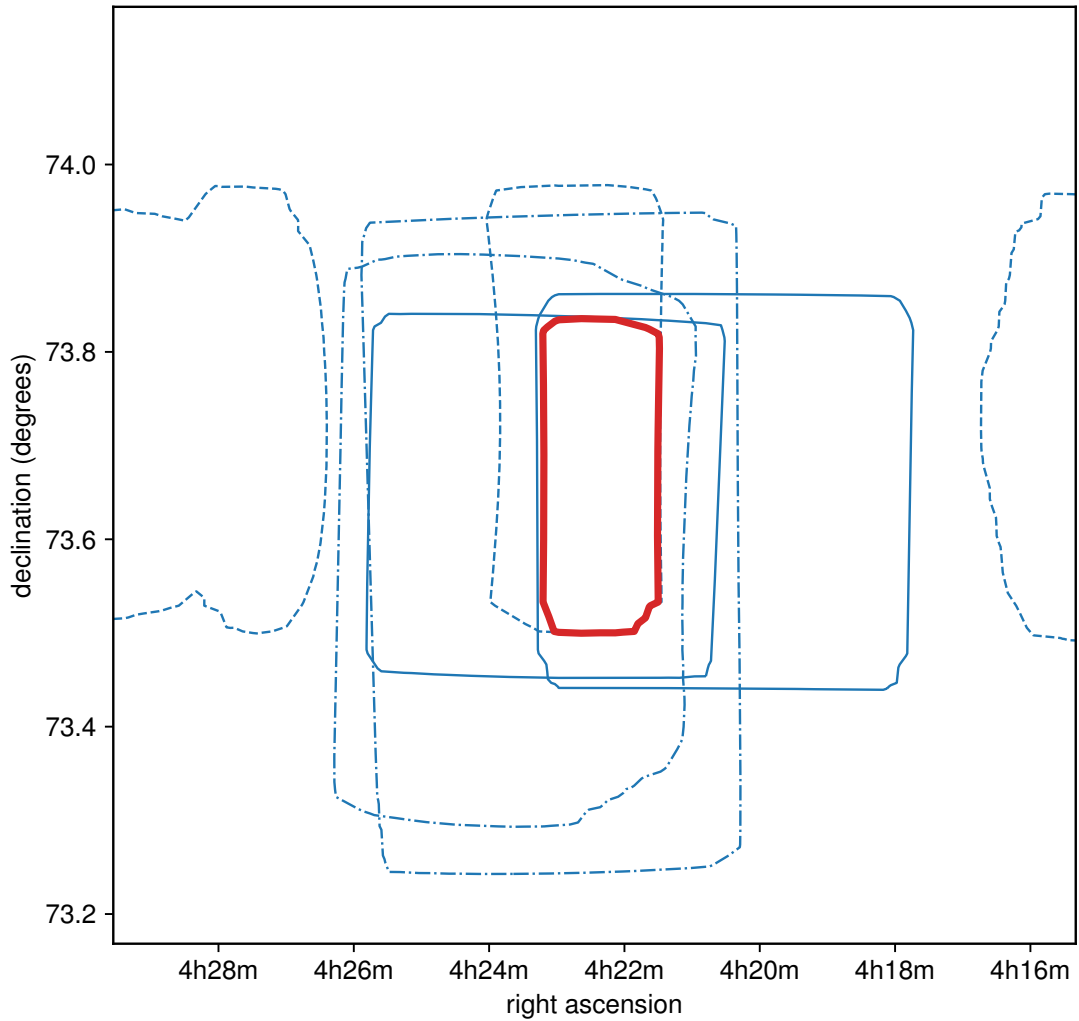

**Figure 1: Localization of FRB 180814.J0422+73.** Contours of positional likelihood (99%) in blue for all five CHIME/FRB events detected from FRB 180814.J0422+73, based on analytically estimated beam models specified for each detection beam, as well as the combined 99% confidence region in red. The dashed line corresponds to the multi-beam detection, which allows for several, disconnected regions. The dash-dotted lines correspond to the lower anti-podal detections, while the solid blue lines correspond to the remaining single beam detections. The combined position is consistent with none of the five events having originated in a sidelobe. See Methods for details.

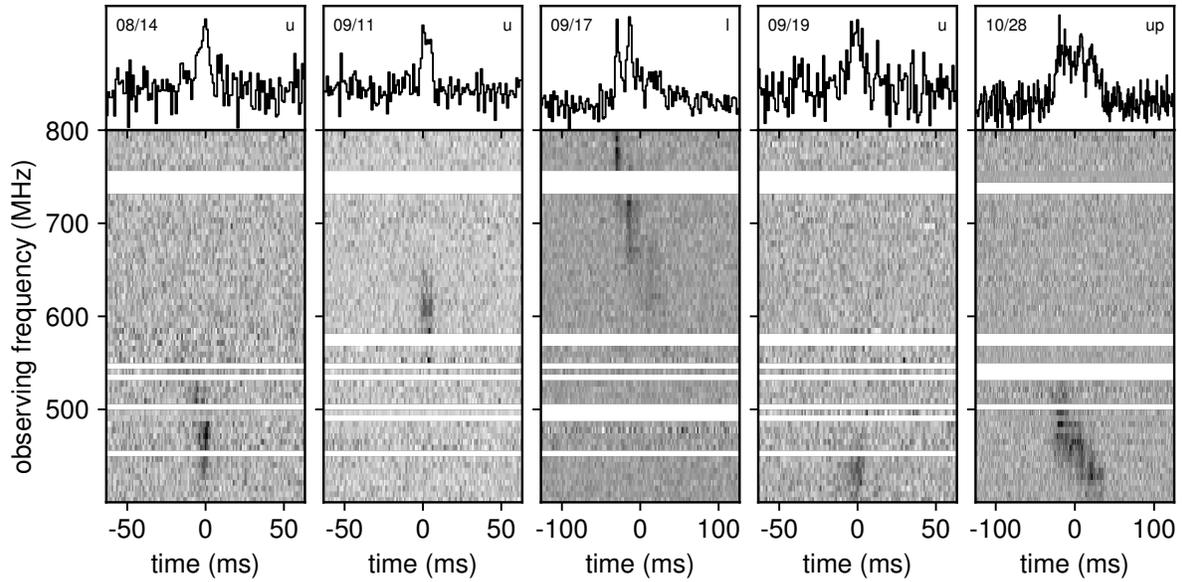

**Figure 2**: **Radio frequency and time ("waterfall") plots for the bursts of FRB 180814.J0422+73.** Four CHIME/FRB-detected events and one CHIME/Pulsar-detected event from FRB 180814.J0422+73 (including the first, also shown in our companion paper[6]) for which raw intensity data were saved. A fifth CHIME/FRB-detected event has metadata only and is not shown. The greyscale intensity is proportional to SNR. The top line plots are frequency-summed burst profiles. Note the differing durations on the time axes. Importantly, the frequency-dependent responses of the primary and formed beams, which can have variations of order unity on ~15-MHz scales, have not been accounted for in these plots. Here, all the bursts were dedispersed with DM = 189.4 pc cm$^{-3}$, a value consistent with that inferred for all when dedispersing to maximize sub-pulse structure (see Table 1). "U" and "L" refer to upper and lower (or anti-podal) transit, respectively. "P" refers to the burst detected with the CHIME/Pulsar backend. The frequency axis has been averaged to 64 frequency channels for display and wide-band channels that are contaminated with persistent RFI are masked and coloured white. Narrow-band regions that are contaminated or that have no phase calibration have been masked and interpolated over for display. After masking, the effective bandwidth ranges from 192 to 267 MHz for these data.

# Methods

**Status of CHIME Telescope and CHIME/FRB Detection Instrument:** The status of the CHIME telescope during July and August 2018 when the first burst from FRB 180814.J0422+73 was detected was described in our companion paper[6]. The remaining bursts were detected in September and October when the telescope configuration was still being modified frequently for commissioning purposes, but which had the following major differences relative to in July and August: (i) the N-S formed beam range was configured from −60º to +60º in zenith angle, providing ~170 deg$^2$ FOV, in contrast to the −90º to +90º extent that provided ~250 deg$^2$ FOV previously; (ii) we had completed installation of all CHIME/FRB compute nodes[7] so generally had four independent E-W beams for each declination. These were spaced 0.4º degrees apart, with one at zenith. (iii) Our calibration method was different, as explained below.

We multiply the time stream from each analog input in the X-engine by a frequency-dependent, complex-valued gain prior to combining time streams into formed beams. This complex gain is calibrated using data from a narrow window around the transit of a radio-bright point source. Cygnus A was used as the calibrator source for the entire period between July and November. The algorithm for determining the complex gains is described in our companion paper[6]. In summary, an eigenvalue decomposition of the $N^2$ visibility matrix is used to estimate the response of each feed to the calibrator. This is done independently for each of the native 1024 frequency channels. The complex gain is then given by the inverse of the point source response after correcting for the contribution to the phase due to the known geometrical delay between feeds. Before September 4 the complex gain calibration was performed roughly once per week. After September 4 we performed complex gain calibration once per night, approximately 1 hour after the transit of Cygnus A.

On September 4 the algorithm was updated to provide amplitude calibration in addition to phase calibration. The amplitude calibration is given by the square root of the spectral flux density of the calibrator divided by the magnitude of the response to the calibrator inferred from the eigenvalue decomposition. The spectral flux densities are obtained from measurements by the Karl G. Jansky Very Large Array (VLA)[29] interpolated to the CHIME band. The amplitude calibration accounts for differences in the antenna gain between the two polarizations that are on average 15%, as well as variations in the antenna/analog gain over feeds of a given polarization that have an RMS of approximately 12%.

Prior to September 4 all analog inputs were included in the construction of formed beams, independent of operational status. On September 4 we began setting the complex gain to zero for 1 − 2% of the analog inputs. This corresponds to a static list of 23 out of 2048 inputs that are either not connected to feeds or have non-operational receiver chains, and a dynamic list of as many as 17 additional inputs that show anomalously low response to the calibrator over a large fraction of the band.

On September 15 the calibration algorithm was updated to improve how the two orthogonal polarizations are treated during the eigenvalue decomposition. For the majority of frequencies this had a negligible impact (< 1%) on our estimates of the complex gain. However, for small groups of frequencies where the magnitude of the response of the two polarizations is comparable (specifically 510–518 MHz, 540–548 MHz, and 600–608 MHz) this update corrected an overestimation of the amplitude of the gain that was as large as a factor of 5.

Starting September 4 the calibration algorithm compared the largest eigenvalue during the transit to the largest eigenvalue from recent off-source data to determine if the calibrator was the dominant signal. If the ratio was less than 3, then the gains were assumed invalid. On average, the calibration algorithm failed to obtain gains for 30% of the frequency channels. These frequencies correspond primarily to regions of the band contaminated by persistent sources of RFI, but also include frequencies with intermittent RFI and frequencies that show a decorrelation due to data frame misalignment. Prior to October 7, we set the gain to zero for all feeds at these frequencies. After October 7 we interpolated the gains from the valid frequencies, so long as the uncertainty on the interpolation was less than 1%. Typically, this process recovered gains for 10% of the frequency channels.

The strategy outlined above effectively calibrates the bandpass for sources at transit at the declination of the calibrator. However, the spectrum of sources at other hour angles and declinations will be modulated by the frequency dependent beam pattern. In this work, we do not attempt to correct for the beam pattern and determine source spectra. We also do not attempt to correct for variations in the complex gain on timescales less than one day. We estimate these daily variations to be less than 3% in amplitude (and primarily common mode over feeds) and less than 0.08 radians in phase.

**Localization of FRB 180814.J0422+73:** With a frequency-dependent sensitivity model of the primary and formed beams, it is possible to constrain the location of a burst using techniques similar to those presented[30] and applied[31] elsewhere. The model is used to compute expected ratios between the SNRs recovered by neighbouring beams for a grid of sky locations and spectral indices. Since the SNR values have unit variance by definition, the comparison of observed SNRs to model-predicted SNRs can be cast as a grid search $\chi^2$-minimization, which provides a natural way to compute uncertainties. Non-detections contribute to the $\chi^2$ value of a grid point only if the expected SNR is above threshold. Marginalizing over spectral index allows multiple bursts to be combined by summing $\chi^2$ values for each position. Detections from different beam configurations can be combined provided the appropriate model is used for each burst. The 99% uncertainty region after combining the five CHIME/FRB bursts is closely approximated by a rectangle of (J2000) R.A. $4^h 22^m 22^s$ with uncertainty ±4′ and Dec. 73°40′ with uncertainty ±10′ (both 99% confidence). The quoted R.A. uncertainty has been scaled by $cos$(Dec.) to reflect the angular extent on the sky. To verify the accuracy of this method, 52 pulsars that primarily produce single-beam detections were selected as analogs. For each source, several randomized samples of five events were used to create joint localizations. 47 pulsars consistently produced localizations where the true position was contained in the 99% uncertainty region, while the five remaining sources are in agreement after considering the large uncertainties in their catalogue positions. We do not identify any significant systematic errors from the residuals and conclude that our current beam model is sufficient for this particular detection scenario.

**Burst Dedispersion and Property Determination:** For dedispersion and characterization of the bursts, we preprocessed the 16,384 frequency channel raw intensity data by (a) masking channels that have missing data, did not have a phase calibration for beamforming, or are known to contain narrow-band RFI (the resulting effective bandwidths used in the analyses of the bursts presented here are 237.5, 192.5, 267.2, 217.6 and 211.0 MHz, respectively), (b) subtracting the time-axis mean and dividing by the time-axis standard deviation for each channel, and (c) sub-banding the data as required by averaging over the unflagged channels. To optimize SNR, we dedispersed the 16,384-channel intensity data to a nominal DM of 189 pc cm$^{-3}$ and sub-banded the data to 1,024 frequency channels. We then dedispersed the data over a range of trial DMs and convolved the summed time series with a range of 1D boxcar kernels that are normalized by the square root of their widths. The resulting SNR vs. DM curve was fit by a Gaussian. We define the mean of the fitted distribution as the DM measurement and the standard deviation as its uncertainty. The resulting DMs for the five bursts are combined to find a weighted mean and uncertainty. To optimize temporal structures in the bursts, we fit for the sharpness of the peaks in the frequency-channel-summed time series. To do so, we calculated the power spectrum of an on- and an off-pulse window, knowing that sharper peaks need a larger number of frequencies to resolve the edges. We use the index of the first frequency in the on-pulse power spectrum for which the amplitude is less than 3 times the RMS of the off-pulse power spectrum as a metric for the sharpness of the peaks and we attempt to maximize this metric as a function of DM. We fit a Gaussian to the resulting sharpness metric vs. DM curve, where we call the mean and standard deviation the structure-optimizing DM measurement and corresponding uncertainty.

In order to quantify sub-pulse structure in the September 17 CHIME/FRB and the October 28 CHIME/Pulsar burst, we model three/five components by three/five 2D Gaussians. After binning the 600–800/400–600 MHz intensity data to 64 frequency channels and masking out bad channels, we fit the peak locations, standard deviations, amplitudes and a background offset, using least-squares optimization, leaving the DM fixed to 189.4 pc cm$^{-3}$. Extended Data Table 1 shows the best-fit parameters and Extended Data Fig. 1 shows the data, best-fit models and residuals. This analysis is affected by our frequency-dependent sensitivity which has order-unity structure on 15 MHz scales due to the beam response. To quantify this, we extract the frequency-dependent sensitivity for a grid of beam positions at the time of the September 17 burst in the uncertainty region around the best-known sky position of FRB 180814.J0422+73. We consider the beam position in the uncertainty region that would lead to the largest variation over the receiver bandwidth and we repeat the sub-pulse fitting analysis after applying the sensitivity correction. We find that all peaks move down in frequency by at most ~10 MHz and use this as a rough estimate of the systematic error in the centre frequencies of the components, which we add to the statistical error in quadrature for subsequent analysis. We fit linear models through the component peaks and find frequency drift rates of −6.4±0.7 and −1.3±0.3 MHz ms$^{-1}$.

**CHIME/Pulsar Instrument and Data Analysis:** The CHIME X-engine is also designed to form 10 independent tracking coherent beams by spatially correlating all the 1024 receiver signals phased to specific celestial positions. The resultant data streams are complex voltages, retaining the full time resolution and phase information. These "tracking" formed beams are useful for observing pulsars or other known radio transients for extended periods of time as they transit the CHIME primary beam.

A separate backend for pulsar-timing/filterbank observations has been constructed to process the tracking-beam streams and is currently under commissioning. The CHIME/Pulsar backend[32] consists of ten compute nodes, each node receives one beamformed stream of complex voltages. In the chosen CHIME/Pulsar configuration, the node records polarization-summed coherently dedispersed filterbank data (i.e., time series of dynamic spectra) at 327.6 μs, a factor of 3 higher than the CHIME/FRB intensity stream, while the original frequency resolution from the F-engine is preserved. A detailed description of the CHIME/Pulsar instrument will be presented elsewhere.

For the work described here, the CHIME/Pulsar backend monitored the field of FRB 180814.J0422+73 daily starting October 26, each time recording polarization-summed filterbank data which was coherently dedispersed at a DM of 189 pc cm$^{-3}$. From October 26 to November 4 we obtained 9 hours (upper transit) and 2.4 hours (lower transit) of exposure. These data were examined for single pulses using the presto[33] software suite which resulted in the detection of the October 28 event in Table 1 and shown in Fig. 2.

**Exposure Time Estimation:** We observed the location of FRB 180814.J0422+73 with the CHIME/FRB system for 23 hours in total, split between 14 and 9 hours for the upper and lower transit, respectively. The reported time is calculated by adding up the duration of daily transits of the source from July 25 through November 4. For the purpose of the calculation, we consider the source to be transiting if it is located in the FWHM region of any of the FFT-formed beams and data from the particular formed beam is being sent to the CHIME/FRB pipeline. Additionally, we only include transits for which the pipeline was fully operational with the raw intensity data being buffered to disk. The FWHM region for the formed beams is calculated for a frequency of 600 MHz using the analytic model for FFT beamforming[7,11] and extends approximately 0.3º in the E-W direction and 0.3º and 0.6º in the N-S direction, for the upper and lower transit, respectively. A different beam model was used for this calculation for the pre-commissioning and commissioning phases owing to the changes in E-W spacing and N-S extent of the beams.

The system sensitivity showed large variations during these 23 hours of observing time. To characterize CHIME/FRB's sensitivity during the upper transit, we identified 5 bright pulsars (PSRs B0105+65, B2224+65, B2241+69, J0737+69 and J1647+66) with 65º < Dec. < 80º that are detected regularly by CHIME/FRB since the start of observations. We consider the median SNR of pulsar detections in the centres of the formed beams as a proxy for the telescope's sensitivity at that time. Hence, for every day that we include in the exposure time to FRB 180814.J0422+73 where we detect at least two pulsar pulses above threshold, we calculated the median SNR, corrected for detection pulse width, of all pulses from each pulsar detected within 10′ from the centres of the formed FFT-beams. We then convert this value into a sensitivity using the radiometer equation to find $T_{sys}/(G\Delta v)$ for each pulsar for each day, where $T_{sys}$ is the system temperature, $G$ the receiver gain and $\Delta v$ the receiver bandwidth. We used catalogued pulsar parameters converted to peak flux density at 600 MHz as a reference[34,35]. Finally, we combine all pulsar transits on one day by calculating the median value and standard deviation and we convert that number in a minimal detectable fluence, using SNR=8 as detectability threshold. We find the median of the minimum detectable fluence over all days included in the reported observing time to be 13±8 Jy ms, for a pulse width of 7.9 ms corresponding to the first burst detected from this source. Although there are sizable uncertainties, we find this estimate to be consistent with the fluence estimated for the burst (see Table 1) using transits of bright sources with well known fluxes such as Cassiopeia A and Cygnus A.

To estimate the reduction in primary beam response between the source's upper and lower transit, we use a radio galaxy ICRF J145907.5+714019 which is located within a 2º declination range of the position of FRB 180814.J0422+73 and has a flux density of 12 Jy. We use the set of $N^2$ visibilities for both transits of this source and apply the beamforming process used for the CHIME/FRB system. By fitting a Gaussian and a polynomial background model to each of the 1024 frequency channels, we find the ratio of the peak response between lower and upper transit to have a median value of ~25%, with a strong dependence on frequency. We note that this analysis has been done for a single data set and the unstated uncertainty in the reported loss in sensitivity for the lower transit could be substantial due to strong zenith angle dependence of the primary beam. Therefore, we only use the bursts detected in the upper transit with the CHIME/FRB system to estimate a 95% confidence lower limit on the burst rate of 0.09 bursts per hour.

**Optical and Infrared Surveys:** The PanSTARRs catalog[24] has 1623 sources in the 99% localization region to a depth of *g, r, i, z, y* < 23.3, 23.2, 23.1, 22.3, 21.3, respectively. Most of these (≈ 70%) are stars based on the difference between the point spread function fitted magnitude and the Kron aperture magnitudes[36]. However, we do not have sufficient information about the remaining ≈ 450 objects to characterize them. The brightest galaxy in the 99% localization region has a brightness of $m_r$ =18.5 AB mag and has large chance coincidence probability and does not constrain the host candidate.

**The radio source NVSS J042149+734107:** NVSS J042149+734107, has an NVSS peak flux of 6.4 mJy beam$^{-1}$ (7.6 mJy integrated flux), but has an upper limit of 0.33 mJy beam$^{-1}$ in the VLASS quicklook images[25]. Due to the large difference in survey resolutions (45″ for NVSS, 25″ for TGSS, and 2″.5 for VLASS), it is possible that the source was resolved in VLASS and hence is undetectable. The VLASS data were acquired in the B-configuration of the JVLA with an outer angular scale of 58″ at 3 GHz. We smoothed the VLASS quicklook images at angular scales of 5″ – 60″ in steps of 5″. At a scale of 10″, there is some possible faint double-lobed structure, however this location lies on the sidelobe of a bright object, so we cannot confirm whether the double-lobed structure is real. On the other hand, if this source is unresolved in VLASS, for a typical spectral index $\alpha = -0.7$, the source must have faded by a factor of > 11. Conversely, if the source has not varied, it would have an unusually steep $\alpha < -3.9$. NVSS J042149+734107 is also not detected in 150-MHz Tata Institute of Fundamental Research Giant Meterwave Radio Telescope Sky Survey (TGSS)[37] that observed in 2010-2012 to an RMS depth of 5 mJy beam$^{-1}$. This restricts the spectral index $\alpha > -0.57$.

**Estimation of Event Fluences:** Fluence estimation methods are presented in the companion paper[6] and summarized here. We assumed North-South beam symmetry and that the bursts were detected in the centre of the FFT-formed beams. We used observations on November 18 of three bright sources with well known fluxes, Tau A, NGC 7720, and 3C 133, to obtain a flux conversion versus frequency in the approximate direction (within 5º elevation angle) of the events to account for the telescope primary beam. The resulting calibration was applied to the spectrum of the events and fluences were calculated.

We also assessed the time variation of the flux conversion by observing these sources over several days, and included this RMS variation in the reported fluence error. We used the bright sources to assess fluence error by applying the calibration of each of the point sources to estimate the flux of the others.

For this commissioning data, a large fraction of the bandwidth was flagged and masked when the events were triggered due to frequency channels that were contaminated with persistent RFI (about 20-25%), correlator processor nodes that were offline, or a failure in the calculation of phase calibrations for certain frequency channels in the nightly calibration needed as input to the beam former. The recorded bandwidth for these events ranged from 190–239 MHz. In order to apply the flux conversion factor, additional frequency channels that were unusable for the bright source observations resulted in a further reduction of bandwidth by 10–23 MHz. The remaining bandwidth was used for the fluence estimates and corresponds to 216, 204, 178, 200 MHz for events 180814, 180911, 180917, and 180919, respectively. No fluence measurement is possible for event 180906 for which we did not record intensity information, and no measurement is presented for 181028 as the CHIME/Pulsar instrument flux conversion factors have not yet been cross-calibrated with CHIME/FRB.

We note that we do not expect the reported SNR to be linearly proportional to fluence because (1) the intensity spectrum used for the SNR is not calibrated for the frequency-dependent response on the primary and formed beams, and (2) the primary beam gain differs by a factor of four between the upper and lower transits. The first effect can be substantial, particularly in the presence of narrow spectral structure.

**Data availability** The data used in this publication are available at https://chime-frb-open-data.github.io/

**Code availability** The code used for our data analysis is available at https://chime-frb-open-data.github.io/

| Burst | Component | Time (ms) | Width (ms) | Center frequency (MHz) | Bandwidth (MHz) |
|---|---|---|---|---|---|
| 09/17 | 1 | 0.0(1) | 3.0(2) | 775(4) | 83(9) |
|  | 2 | 8.1(2) | 6.8(5) | 711(3) | 85(6) |
|  | 3 | 21.4(7) | 12(2) | 636(8) | 95(20) |
| 10/28 | 1 | 0.0(5) | 11.8(9) | 485(1) | 55(3) |
|  | 2 | 13.0(6) | 12(2) | 468(2) | 54(4) |
|  | 3 | 25.4(3) | 8.0(7) | 455(2) | 58(4) |
|  | 4 | 38.5(3) | 10.2(8) | 433(1) | 45(3) |
|  | 5 | 51.0(4) | 6.6(9) | 422(2) | 34(5) |

Extended Data Table 1: **Sub-pulse parameters.** Best-fit parameters for 3 (5) Gaussian components in the September 17 (October 28) burst from FRB 180814.J0422+73. Width and bandwidth are measured at full-width-half-maximum and uncertainties are statistical one standard deviation. Since the spectrum is modulated by the beam response, which has order-unity structure on 15 MHz scales, centre frequency and bandwidth measurements carry an additional systematic error of up to 10 MHz (see text).

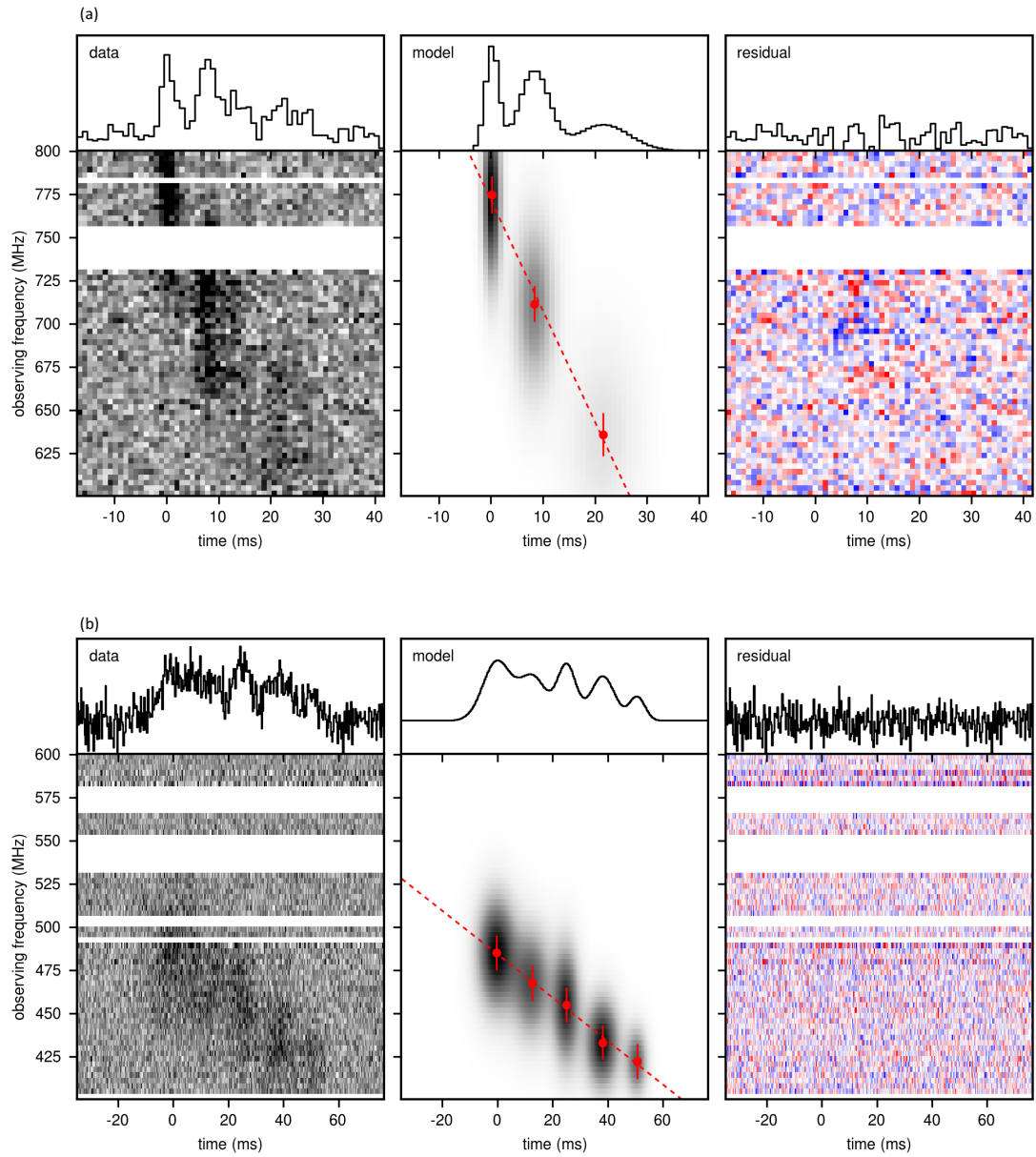

Extended Data Figure 1: **Sub-pulse frequency drift rates.** Sub-pulse model fits for the (a) September 17 CHIME/FRB burst and the (b) October 28 CHIME/Pulsar burst. Dedispersed intensity data (DM=189.4 pc cm$^{-3}$), the best-fit model and residuals, as well as the summed time series are shown. Only the half of the receiver bandwidth in which the burst was detected is used in the analysis. The colour scale for the intensity data and residuals is clipped to ±3$\sigma$ from the median of the residual data and a divergent rather than a sequential colour scale is used for the residuals to guide the eye. Red points overlaid on the models show the centre frequency and 1$\sigma$ statistical uncertainty with a 10-MHz systematic error added in quadrature. The red dashed lines show linear drift rates of −6.4 and −1.3 MHz ms$^{-1}$.